\documentclass[aps,prl,twocolumn,showpacs,superscriptaddress,groupedaddress]{revtex4}  
\usepackage{graphicx}  
\usepackage{dcolumn}   
\usepackage{bm}        
\usepackage{amssymb}   
\usepackage{amsmath} 
 \begin{document}
 	
 	\title{Symmetry Obstruction to Fermi Liquid Behavior in the Unitary Limit}
 	
 	\def\CMU{Department of Physics, Carnegie Mellon University, Pittsburgh, Pennsylvania, 15213, USA}
 	
 	\author{Ira Z. Rothstein}
 	\email[E-mail:]{izr@andrew.cmu.edu}
 	\affiliation{\CMU}
 	
 	\author{Prashant Shrivastava}
 	\email[E-mail:]{prashans@andrew.cmu.edu}
 	\affiliation{\CMU}  
 	\begin{abstract}
	We show that a Fermi gas, in three dimensions, at temperatures above the  superconducting phase transition but below the Fermi temperature, can not be described by Fermi Liquid Theory (FLT) in the unitary limit where the scattering length diverges. The result follows by showing that the there are no effective field theory  descriptions that both behave like a Fermi liquid and properly non-linearly realize the spontaneously broken boost and conformal invariance of the system. We have also derived an exact result for the beta function
for the coupling function in the unitary limit. 
 	\end{abstract}
 	\maketitle
 	\section{Introduction}
 	Landau's Fermi liquid theory (FLT) is a theory of interacting fermions which describes the normal state of metals \cite{Landau,kohn,Baym}. In modern parlance the ubiquity of Fermi liquid behavior is a consequence of the fact that under a certain set of generic assumptions the long distance or low energy behavior is governed by a universal fixed point.
	That is, Fermi liquids fall into a universality class. The starting assumption of the theory is that the interacting system
	 can be reached by beginning with a free theory and adiabatically turning on the interactions such that the free fermion states  evolve into  interacting quasi-particles with the same charge and spin as an electron but not necessarily the same mass. Landau showed that, under these assumptions, the width ($\Gamma$) of these quasi-particles  is suppressed due to Pauli blocking of the final states such that
 	$ 	\Gamma(E)\sim{(E-\mu)^2},  
 	$
 	where $\mu$ is the chemical potential. This \textit{a posteriori} justifies the notion of a quasi-particle. The signature of FLT behavior in the normal phase of metals has generic features such as a resistivity scaling as $T^2$, the existence of zero sound and long lived gapless excitations. 
	
	If the theory is weakly coupled at energies well above $E_F$, there is good reason to believe that Fermi liquid behavior will arise at long distances, whereas for strongly interacting theories it need not be the case, and non-Fermi liquid behavior, as in high $T_c$ compounds,  arises such that the  gap, which appears below $T_c$, leaves vestiges,  so-called   ``pseudo-gaps'' (for a discussion see \cite{mueller1}), as the temperature is raised above $T_c$.  This has raised the question as to whether or not pseudo-gaps are generic features of strongly coupled systems of fermions.
 
 A prime laboratory for such ``strange metal'' behavior is systems  of cold atoms, which when sufficiently dilute, mimics fermionic many body systems.
 The utility of these systems is that the scattering length  ($a$) can be tuned. 
When $a>0$ the systems behaves like a Fermi liquid with a BCS like phase transition involving the condensation of weakly bound fermions (Cooper pairs). As the coupling is  increased,  $a$ grows and eventually changes sign at which point the system behaves more like a Bose condensate (BEC) with  pairs of atoms forming bosonic molecules and condensing. The in-between point where $1/a \rightarrow 0$ is called the ``unitarity limit'' (\cite{levin}). 

There is no consensus in the literature as to whether such systems at unitarity  behave like Fermi liquids. Ref.(\cite{natureFLT}, \cite{PRLFLT}) found Fermi liquid behavior above $T_c$ while (\cite{psuedogap1},\cite{psuedogap2},\cite{levin2}) did not. It is important to take into account the fact these experiments were measuring different observables and
the conclusions may assume that pseudo-gap behavior can not mimic Fermi liquid behavior for some sub-set \cite{mueller2} of these observables.
 	
It is a theoretical challenge to understand such strongly coupled systems analytically as there is no small expansion parameter. Nonetheless,  a tremendous amount can be learned about strongly coupled systems when
symmetry considerations are taken into account. Given that the unitarity limit is a point of enhanced (conformal) symmetry one might hope for an increase in predictive power in this regime. In this letter we will show that this is indeed the case and that it can be shown from first principles that in three dimensions Fermi gases in the unitary limit can not be described by FLT when $T_F>T>T_c$, where $T_c$ is the critical temperature for the superfluid phase transition and $T_F$ is the Fermi temperature.

 	The starting point of our analysis is the effective field theory (EFT) of Fermi liquids developed in Ref.~\cite{benfatto,shankar, polchinski} which  is based on an expansion around the Fermi surface and becomes exact in the infra red limit where $E /E_F \rightarrow 0$. This is the unique EFT that describes the universality class that is FLT. The founding assumptions is the existence of long lived  quasi-particles with the quantum numbers of the electron.  Once the theory has been defined, one shows that it does predict a quasi-particle width, $\Gamma(E) \sim (E-\mu)^2$, a self-consistency check and a defining characteristic of FLT.
The crux of our argument is that at unitary this consistency check fails thus eliminating the possibility of standard FLT  behavior in this limit.	
	
	 Using power counting arguments, it can be shown that the only relevant interactions near the Fermi surface are, the forward scattering (FS) of quasi-particles or the superconducting (BCS) back to back interaction between quasi-particles\cite{ benfatto,shankar,polchinski}. The same conclusion can also be reached by using only the RG invariance and the Galilean boost symmetry  \cite{us}. As such, the Fermi liquid action can be written as
 	\begin{align}
 	\label{FL} 
 	S&=\int dtd^dp~\psi^{\dagger}_{\vec{p}}(t)\left(i\partial_{t}-\varepsilon(\vec{p})+\mu\right)\psi_{\vec{p}}(t)-\frac{1}{2}\prod_{i=1}^{4}\int dtd^dp_{i} \nonumber \\ & \delta^{d}\left(\sum_{i=1}^{4}\vec{p}_{i}\right)g(\vec{p}_1,\vec{p}_2,\vec{p}_3,\vec{p}_4)\psi^{\dagger}_{\vec{p}_1}(t)\psi^{\dagger}_{\vec{p}_2}(t)\psi_{\vec{p}_3}(t)\psi_{\vec{p}_4}(t)
 	\end{align}
 	where the coupling function $g$ is restricted to  forward scattering ($g_{FS}$) or a BCS back-to-back ($g_{BCS}$) kinematic configurations.  The energy functional $\varepsilon(p)$ leads to a generalized dispersion relation $E= \epsilon(p)$  which when expanded around a point ($\vec p_F(\theta)$)  on the Fermi surface, $\vec p = \vec p_F(\theta)+\delta \vec k$, gives  $\epsilon(p_F(\theta)) \approx \vec v_F(\theta) \cdot \delta\vec k$, with $v_F$ being the fermi velocity.
	The BCS coupling (in some channel) grows in the IR and leads to condensation of Cooper pairs, whereas the
	forward scattering interaction is RG invariant by power counting arguments (see e.g. \cite{shankar}).

A crucial part of our analysis relies on insisting that the low energy theory properly realizes the space-time and internal
symmetries of the short distance physics which are Galilean invariance and particle number conservation, the latter of which, along with the translational sub-group \footnote{In a metal of course this is broken and the arguments here change slightly \cite{us}.},  are explicitly  realized in Eq.~(\ref{FL}). Rotational invariance implies that the Fermi velocity is a constant  and the coupling function $g(p_i)$, is only a function of the relative angle between the 3-momentum vectors. The only spontaneously broken symmetry\footnote{Away from unitarity.} is 
the Galilean boost invariance which is not manifest in Eq.~(\ref{FL}).

\section{Three Paths to Symmetry Realization}

 	Naively the spontaneous breaking of  the Galilean boost symmetry should, by Goldstone's theorem \cite{Goldstone:1961eq,Goldstone:1962es}, lead to the  existence of a massless scalar boson called the \textit{framon} \cite{framids}. However, 
as is well known, when space-time symmetries are broken there need not be a one-to-one map between broken generators
and Goldstone bosons. This is usually explained as being due to an ``inverse Higgs constraint" (IHC)\cite{Ivanov,Volkov,Ogievetsky}.
These constraints arise as a consequence of the fact that it is often possible that only one Goldstone boson is needed
to assure invariance under multiple symmetry transformations \cite{ML}.
  The conditions for the existence of an IHC are \cite{Ivanov,Volkov,Ogievetsky}
 	\begin{equation}
 	[P_{\nu},X']~\supseteq~X,
 	\end{equation} 	
	 where  ($P_{\nu}$) is an unbroken translation (which may include internal translation (see for instance \cite{framids})  and  $X^{\prime}$ and $X$  are broken generators.
When this condition is met one can eliminate the Goldstone boson for $X^{\prime}$ ($\pi^{\prime}$) in favor of  $X$ ($\pi$). This is accomplished by setting, 
$\nabla_{\nu}\pi$ (to be defined below) to zero,  which results in an algebraic relation between $\pi$ and $\pi^{\prime}$.
 A classic example of this is a crystal where there are no independent Goldstone bosons 
for the broken rotations, as phonons suffice to saturate all the Ward identities. 

We would like to point out that it is not necessary to impose an IHC. A theory involving all the Goldstone modes is perfectly acceptable although in practice the theory might be cumbersome to use. This however would be the most straight forward way of realizing the broken symmetries. The second option, which is usually the one used, is to impose all possible IHCs and work with a minimal set of Goldstones. Finally, there is a third possibility
which we call the Dynamical Inverse Higgs Constraints (DIHC) \cite{us}, whereby an operator constraint 
ensures the symmetry is realized. The canonical example of DIHC arises in the case of FLT, where only
boost invariance is broken but there is no possibility for an IHC.

Of the three paths to symmetry, the first two may not be compatible with FLT behavior.
%
%
To see this one first notes
 that when a spacetime symmetry is broken  the Goldstone mode may  be non-derivatively coupled (for a proof of this see \cite{VW,us}), and thus will not necessarily be irrelevant
 in the low energy limit. A simple one loop calculation shows that  \textit{framons} fluctuations generate a quasi-particle width which scales as 
 	\begin{equation}
 	\label{selfenergyQP} 
 	\Gamma(E)\sim (E-\mu)^{d/3}, 
 	\end{equation}  
 	which leads to Non-Fermi liquid behavior in d=2 and Marginal Fermi liquids in d=3. 

As such it is natural to ask how FLT is consistent with boost invariance?
 The answer follows from the process of  elimination, as the only remaining possibility is the existence of a DIHC, the third alternative discussed above.
  In fact, in his seminal work \cite{Landau} Landau showed that
boost invariance implies that the following relation must hold
 	\begin{equation}
 	\label{LR}
 	\frac{1}{m^{\star}}=\frac{1}{m}+\frac{G_1}{3}.
 	\end{equation}
 	Here $m^{\star}$ is the effective mass of the quasi-particles defined by $m^{\star}  v_F = k_F$, $m$ is the free electron mass and $G_1=g_{1}D(\mu)$, where the (FS) coupling function is expanded in Legendre polynomials, $g(\theta)=\sum_{l}g_{l}P_{l}(\cos(\theta))$ and $D(\mu)$ is the density of states on the Fermi surface.  Below we will show how this relation arises due to  a DIHC.

It is interesting to note that a similar situation arises 
when rotational invariance is spontaneously broken by the Fermi surface but translations are unbroken as in case of a nematic Fermi fluid.
Again there are two possible realizations, since there is no IHC. Either a condition similar to (\ref{LR}) is obeyed such that there exists a DIHC for the rotational symmetry, or a collective mode, the ``\textit{angulon}'', must arise in the spectrum which couples non-derivatively and leads to  non-FLT behavior \cite{VW}.
In Ref.~\cite{OKF}, it was shown that  this theory does generate a collective gapless mode that  couples 
non-derivatively to the quasi-particles. This theory does not have any additional Goldstones (i.e. no framon) and so
the boost invariance is only manifest once the Landau relation is obeyed. That is, the rotational symmetry is
realized via a Goldstone mode, but the boost is realized via the existence of a DIHC.
 \section{Fermions at Unitarity}
 	Here we are interested in fermions at unitarity whose short distance effective action is invariant under
	the larger Schrodinger group, which has two added symmetries beyond the standard Fermi liquid namely, dilations and special conformal transformations (SCT). These additional symmetries are also spontaneously broken by the Fermi surface.
	Our goal is to determine whether or not nature can realize these broken symmetries and still retain
	Fermi liquid behavior, and we shall now see that the answer is no in three spatial dimensions.
	
In \cite{oz} it was pointed out that there is no boost invariant kinetic term for the  \textit{dilaton}, the Goldstone associated with the breaking of dilations,  which might explain its apparent absence in nature.  
 However, the existence of \textit{framon}  resolves this issue \cite{us}, as the non-invariance of the naive kinetic term is compensated by a shift in the \textit{framon} field. Hence at unitarity, one possible realization of the symmetries involves three  gapless modes, corresponding to the three (two of which are conformal) broken symmetries.   
 
  The other possibilities are  that the systems realizes the symmetry with fewer  Goldstone bosons due to an IHC and/or a DIHC. 
We will explore both the possibilities in  two and three dimensions. While it is well known that the two dimensional case should behave as a free Fermi gas \cite{NN}, it is instructive to see how this result follows from symmetry requirements.
 
 	
 	\section{Non-linear realizations of spacetime symmetries}
 	To understand how Goldstones realize the spontaneously broken space-time symmetries, we will utilize the
	coset construction  \cite{CCWZ} as applied to space-time symmetries \cite{Ivanov,Volkov,Ogievetsky}.
	 A Fermi liquid tuned to unitarity spontaneously breaks the full Schrödinger group, $\mathcal{G}$ into an unbroken sub-group, $\mathcal{H}$ consisting of the symmetry generators for translations $(H,\vec{P})$, rotations ($\vec{J}$) and $U(1)$ particle number ($M$). The broken generators are boosts ($\vec{K}$), dilations ($D$) and SCT ($C$). An element of the coset space, $\mathcal{G}/\mathcal{H}$ can be parametrized as \cite{Ivanov,Volkov,Ogievetsky},
 	\begin{equation}
 	\label{coset}
 	\Omega=e^{iHt}e^{-i\vec{P}\cdot\vec{x}}e^{-i\vec{K}\cdot\vec{\eta}(x)}e^{-iC\Lambda(x)}e^{-iD\phi(x)},
 	\end{equation}
 	where $\vec{\eta}(x)$ is the \textit{framon}, $\phi(x)$ is the \textit{dilaton} and $\Lambda(x)$ is the Goldstone boson for SCT. Building blocks for writing down $\mathcal{G}$-invariant actions can be obtained from the Maurer-Cartan form via the identification,
 	\begin{equation}
 	\label{MC}
 	\Omega^{-1}\partial_{\mu}\Omega=iE_{\mu}^{\nu}\left(P_{\nu}+ \nabla_{\nu}\vec{\eta}\cdot\vec{K}+D\nabla_{\nu}\phi+C\nabla_{\nu}\Lambda+A_{\nu}M\right).
 	\end{equation}
 	We can use the covariant derivatives ($\nabla_{\nu}\phi$, $\nabla_{\nu}\vec{\eta}$, $\nabla_{\nu}\Lambda$), the vielbein ($E_{\mu}^{\nu}$) and the gauge field ($A_{\nu}$), to write down terms which are linearly $\mathcal{H}$-invariant. 
	Such terms are automatically invariant under the full group $\mathcal{G}$. For a detailed calculation of these covariants derivatives and their coupling to quasi-particles see Ref.~\cite{us}.

 	For the broken generators considered in our set up, we have the commutation relation $[H,C]=iD$ which gives rise to an IHC.
	Using (\ref{MC}) we find
	\begin{equation}
 	\label{cov1}
 	 \nabla_{0}\phi = \Lambda+\partial_{t}\phi+...
 	\end{equation}
	and by setting this result to zero, we may eliminate $\Lambda$ (to lowest order in derivatives) in favor of the \textit{dilaton}.
	Another IHC arises as a consequence of  $[P_{i},C]=-iK_{i}$ via the covariant derivative 
	 	\begin{equation}
 	\label{cov2}
 	\vec{\nabla}\cdot\vec{\eta}=3\Lambda+\vec{\partial}\cdot\vec{\eta}+....
 	\end{equation}
Setting these covariant derivatives to zero (i.e. imposing the IHC) is consistent with the symmetries.
Nature may or may not choose to impose any or all of these constraints.
The final possibility is that there are no Goldstone modes at all due to DIHCs. To understand how this scenario can arise let us first discuss how the DIHC is generated in the canonical Fermi liquid.



 	\section{How do Dynamical Inverse Higgs Constraints Arise?}
	
We begin by studying  the  Fermi liquid away from unitarity.  	The \textit{framon} both acts as a gauge field ($A_{\mu}$) and shows up in the vielbein in the coset construction in Eq.~(\ref{MC}) and its coupling to the quasi-particles is given by replacing the normal derivatives in eq. (\ref{FL}) with the covariant derivatives \cite{us},
 	\begin{eqnarray}
 	\label{QSetacoupling}
 	\nabla_{t}\psi&=&\partial_{t}\psi+\vec{\eta}\cdot\vec{\partial}\psi+\frac{i}{2}M\vec{\eta}^2\psi \nonumber \\
 	\nabla^{i}\psi&=&\partial^{i}\psi-iM\eta^i\psi.
 	\end{eqnarray}
 	According to the standard power counting arguments close to the Fermi surface \cite{polchinski,shankar}, the energy ($\varepsilon(p)-\mu$) and  the component of quasi-particle momentum normal to Fermi surface ($l$) scale as $\lambda$, where $\lambda$ is the rescaling parameter of the RG transformation and $\lambda\rightarrow{0}$ as we approach the Fermi surface. Hence, the symmetry  implies that the scaling of $\eta$ is  the same as $l$ i.e $\eta\sim\lambda$.
	Any other choice would lead to a non-invariant action.
We can deduce the scaling of $\eta$ momentum by using the canonical commutation relation,
 	\begin{equation}
 	\label{CCR}
 	[\eta^i(x),\dot{\eta}^j(x')]=\delta^d(x-x')\delta^{ij}.
 	\end{equation}
 	
 	Since the \textit{framon} has the standard bosonic dispersion relation, $E\sim k$, if we assume $k_{\eta}\sim\lambda^{n}$, then from Eq.(\ref{CCR}) we get,
 	\begin{equation}
 	\label{etascaling}
 	n=\frac{2}{d-1}.
 	\end{equation}
 	In $d=2$, the $k_{\eta}\sim\lambda^2$ which is sub-leading to the quasi-particle momentum and so the $\eta(x)$ field must  be multipole expanded \cite{GR} for the power counting to be consistent.  This implies that the $\eta$ field has no dynamics (the kinetic contribution to the action is ignored) and  at leading order   acts as an auxiliary field  which can be removed from the theory by using leading order equations of motion. Doing so gives the following operator constraint (for details see \cite{us}),  	
 	\begin{align}
 	\label{GL}
 	O^B_i&=\Big(\int \frac{d^dp}{(2\pi)^d}~\psi^{\dagger}_{p} (p_{i}-m \frac{\partial \varepsilon_p}{\partial p_i} )\psi_{p}
 	-\frac{m}{2}\int\prod_{a=1}^4\frac{d^{d}p_{a}}{(2\pi)^{d}}\nonumber \\&\delta^{(d)}(p_{1}+p_{2}-p_{3}-p_{4})\Big(\sum_{i}\frac{\partial{g(p_{a})}}{\partial{p_{i,a}}}\Big)\psi^\dagger_{p_4}\psi^{\dagger}_{p_3}\psi_{p_2}\psi_{p_1}\Big)=0.
 	\end{align}
	
This constraint ensures that the symmetries are properly  realized (without the aid of any FLT behavior violating Goldstone bosons) in both two and three dimensions. 	
Note this relation relates a quadratic to a quartic operator, and imposes strong constraints on dynamics.
Indeed, as is shown in \cite{us}, this constraint,  is sufficient to restrict all
interactions near the Fermi surface to be either forward scattering or back to back (BCS) channel. In this way the
universality of FLT should be thought of as a consequence of the symmetry breaking pattern, as is
customary in critical phenomena.	
	
 	Taking one-particle matrix element of this operator  results in a non-trivial relation between the two parameters ($m^*,g_1$) of the theory, which is nothing
	but the Landau relation (\ref{LR}).  
 	
 	In $d=3$   we are no longer forced to perform the multipole expansion, since the \textit{framon} momentum scales in the same way as the quasi-particles' , $k \sim \lambda$. Thus in a canonical FL (i.e. not at unitarity) \textit{a priori} there is no reason why nature could not choose to
	realize the broken boost invariance via the \textit{framon}. However, if this were indeed the case	
	then coupling of $\eta$  to the quasi-particles would give rise to marginal interactions and result in Marginal Fermi liquid behavior.

	The alternative realization of the theory is trivial to find via a DIHC, since the same line of reasoning holds as in the two dimensional case, except now ignoring the kinetic  part of the \textit{framon} field is simply a mathematical trick to
	find an invariant action.	
		 
	\section{Going to Unitary Limit}  	
 	Let us now consider the Fermi liquid at unitarity in $d$ dimensions and ask whether or not the symmetries
	can be realized in a way consistent with Fermi liquid theory.
	We may choose to impose the IHCs and eliminate two of the three Goldstones' by setting the covariant derivatives (\ref{cov1},\ref{cov2}) to zero.
	In so doing we generate the relation
	\begin{equation}
	\partial \cdot \eta = \dot \phi.
	\end{equation}
	In the case of two dimensions, this condition implies that not only is the \textit{framon} non-dynamical 
	but so is the \textit{dilaton}. Thus we will generate two DIHCs. While the constraint from boost invariance will
	be identical to ($\ref{GL}$), varying the quasi-particle action with to respect to the \textit{dilaton} leads to the vanishing of the operator   
\begin{eqnarray}
\label{dilaton}
\mathcal{O}_{\phi}&=&\sum_{\vec{\bf k}}\int d^dp dt~~\psi^{\dagger}_{\vec{\bf p}}(t)\left[2 \varepsilon(p)- p^i  \frac{\partial \varepsilon}{\partial p_i} \right]\psi_{\vec{\bf p}}(t) \nonumber \\
&+&\frac{1}{2}\prod_{a=1}^{4}\int d^dp_{a} dt~\left[(2-d)g(\vec{\bf p}_{i},\mu)-\vec{\bf p}_{i}\cdot\frac{\partial{g(\vec{\bf p}_{i},\mu)}}{\partial\vec{\bf p}_{i}}-\beta(g)\right]\nonumber \\ &\times&\psi^{\dagger}_{\vec{\bf p}_{1}}(t)\psi_{\vec{\bf p}_{2}}(t)\psi^{\dagger}_{\vec{\bf p}_{3}}(t)\psi_{\vec{\bf p}_{4}}(t)=0 
\end{eqnarray}
 	where $\beta(g)=\mu\frac{\partial{g}}{\partial\mu}$ is the beta function.  
	A detailed derivation of this result can be found in \cite{us}. A short path to the derivation follows by calculating
	the Noether charges from the action in (\ref{FL}) for the Schrodinger group and imposing the commutation relation $[D,H]=2iH$, where $D$ is the generator
	of dilatations and $H$ is the Hamiltonian.
	
Let us now see if a Fermi liquid description is consistent with these constraints.
Given our assumption of rotational invariance and the notion of a  well defined Fermi surface, the marginal coupling  is only a function 
of the angles which are scale invariant. Thus the second term in the second line of (\ref{dilaton}) vanishes ($\vec{\bf p}\cdot\frac{\partial{g(\vec{\bf p},\mu)}}{\partial\vec{\bf p}}=0$) , as such if we take
the one particle matrix element we see that the quadratic and quartic terms must vanish
separately since the quadratic term will depend upon the amplitude of the incoming external 
momentum whereas the quartic term will be independent of it due to rotational invariance. This gives the constraint 
\begin{equation}
(2-d)g(\vec{\bf p},\mu)=\beta(g)
\end{equation}
In three dimensions we see that the coupling has power law running which 
is inconsistent with Fermi liquid theory, and in two dimensions the theory is free. Thus we conclude that, if the symmetries are realized without the existence of Goldstones then fermions at unitarity do not behave like Fermi liquid. In two
dimensions the Goldstone can not be dynamical and as such we have ruled out any other
way to realize the symmetry. In three dimensions, where the Goldstone can be dynamical, they lead
to overdamping of quasi-particles, and thus we may conclude that { \it fermions at unitarity can not be be described as a Fermi liquid.} However, our method does
not shed light on the question of whether the breakdown is due to pseudogaps.

We can also consider how these symmetry constraints can be utilized if we assume that the microscopic
theory is defined via the action (\ref{FL}) independent of any Fermi liquid description. Taking the one particle matrix element of
(14)  it is still true, that the quadratic and quartic terms must vanish separately, even if $\vec{\bf p}_{i}\cdot\frac{\partial{g(\vec{\bf p}_{i},\mu)}}{\partial\vec{\bf p}_{i}}\neq 0$, since the quartic term is independent of
the external momenta while the quadratic is not.    
In this case we have the constraints
\begin{eqnarray}
\label{beta2}
\varepsilon &=& \frac{p^2}{2m^\star} \nonumber \\
0&=&(2-d)g(\vec{\bf p}_{i},\mu)-\vec{\bf p}_{i}\cdot\frac{\partial{g(\vec{\bf p}_{i},\mu)}}{\partial{\vec{\bf p}_{i}}} -\beta(g).
\end{eqnarray}
For S-wave scattering ($g(p)$=constant), $m=m^\star$ but for higher angular momentum channels, Eq.(\ref{beta2}) gives us the beta function to all orders.

\section{Conclusions}
In this letter we have addressed the open question  of whether or not three dimensional Fermi gases at unitarity behave like Fermi liquids. We have shown that for temperatures in the range $T_F>T>T_C$ the system will not manifest Fermi liquid behavior.
Our arguments are based solely on the symmetry breaking pattern. In particular, we use the fact that while the conformal
and boost symmetries are spontaneously broken, these symmetries must still be realized, albeit non-linearly, in the low
energy theory. This this can happen either via the existence of the appropriate
Goldstone modes, or through a Dynamical Inverse Higgs Constraint, whereby a non-trivial operator constraint
must be obeyed to manifest the symmetry. We have shown that in three dimensions, neither of these possibilities
is consistent with Fermi liquid behavior.  We have also derived an exact result for the beta function
for the coupling function in the unitary limit.
\section*{Acknowledgements}
This work supported by the DOE contracts DOE DE-FG02-04ER41338 and FG02- 06ER41449. The authors thank Riccardo Penco for comments on the manuscript.

\end{document}